\documentclass[flushrt]{emulateapj}

\usepackage{subfigure} 
\usepackage{epsfig}
\usepackage{color}
\usepackage{tabularx}
\usepackage{booktabs}
\usepackage{multirow}

\newcommand{\npar}{\par \vspace{2.3ex plus 0.3ex minus 0.3ex}}

\usepackage{lscape}
\usepackage{longtable}

\begin{document}
\title{A consistent study of metallicity evolution at $0.8 < z < 2.6$} 

\author{Eva Wuyts\altaffilmark{1}, Jaron Kurk\altaffilmark{1}, Natascha M. F{\"o}rster Schreiber\altaffilmark{1}, Reinhard Genzel\altaffilmark{1,2,3}, Emily Wisnioski\altaffilmark{1}, Kaushala Bandara\altaffilmark{1}, Stijn Wuyts\altaffilmark{1}, Alessandra Beifiori\altaffilmark{1,4}, Ralf Bender\altaffilmark{1,4}, Gabriel B. Brammer\altaffilmark{5}, Andreas Burkert\altaffilmark{4}, Peter Buschkamp\altaffilmark{1}, C. Marcella Carollo\altaffilmark{6}, Jeffrey Chan\altaffilmark{1}, Ric Davies\altaffilmark{1}, Frank Eisenhauer\altaffilmark{1}, Matteo Fossati\altaffilmark{4,1}, Sandesh K. Kulkarni\altaffilmark{1}, Philipp Lang\altaffilmark{1}, Simon J. Lilly\altaffilmark{6}, Dieter Lutz\altaffilmark{1}, Chiara Mancini\altaffilmark{7}, J. Trevor Mendel\altaffilmark{1}, Ivelina G. Momcheva\altaffilmark{8}, Thorsten Naab\altaffilmark{9}, Erica J. Nelson\altaffilmark{8}, Alvio Renzini\altaffilmark{7}, David Rosario\altaffilmark{1}, Roberto P. Saglia\altaffilmark{1,4}, Stella Seitz\altaffilmark{4}, Ray M. Sharples\altaffilmark{10}, Amiel Sternberg\altaffilmark{11}, Sandro Tacchella\altaffilmark{6}, Linda J. Tacconi\altaffilmark{1}, Pieter van Dokkum\altaffilmark{8}, David J. Wilman\altaffilmark{1,4} \footnotemark[*]}
\altaffiltext{1}{Max-Planck-Institut f\"{u}r extraterrestrische Physik, Giessenbachstr.~1, D-85741 Garching, Germany (evawuyts@mpe.mpg.de)}
\altaffiltext{2}{Department of Physics, Le Conte Hall, University of California, 94720 Berkeley, USA}
\altaffiltext{3}{Department of Astronomy, Hearst Field Annex, University of California, Berkeley, 94720, USA}
\altaffiltext{4}{Universit{\"a}ts-Sternwarte M{\"u}nchen, Scheinerstr. 1, M{\"u}nchen, D-81679, Germany}
\altaffiltext{5}{Space Telescope Science Institute, Baltimore, MD 21218, USA}
\altaffiltext{6}{Institute of Astronomy, Department of Physics, Eidgens{\"o}sische Technische Hochschule, ETH Z{\"u}rich, CH-8093, Switzerland}
\altaffiltext{7}{Osservatorio Astronomico di Padova, Vicolo dell'Osservatorio 5, Padova, I-35122, Italy}
\altaffiltext{8}{Department of Astronomy, Yale University, P.O. Box 208101, New Haven, CT 06520-810, USA}
\altaffiltext{9}{Max-Planck Institute for Astrophysics, Karl Schwarzschildstrasse 1, D-85748 Garching, Germany}
\altaffiltext{10}{Department of Physics, Durham University, Science Laboratories, South Road Durham DH1 3LE, UK}
\altaffiltext{11}{School of Physics and Astronomy, Tel Aviv University, Tel Aviv 69978, Israel}

\footnotetext[*]{Based on observations obtained at the Very Large Telescope (VLT) of the European Southern Observatory (ESO), Paranal, Chile (ESO program IDs 073.B-9018, 074.A-9011, 075.A-0466, 076.A-0527, 078.A-0660, 079.A-0341, 080.A-0330, 080.A-0339, 080.A-0635, 081.A-0672, 082.A-0396, 083.A-0781, 087.A-0081, 088.A-0202, 088.A-0209, 091.A-0126, 092.A-0082, 092.A-0091) and at the Large Binocular Telescope (LBT) on Mt. Graham in Arizona. This work is further based on observations taken by the 3D-HST Treasury Program (GO 12177 and 12328) with the NASA/ESA Hubble Space Telescope, which is operated by the Association of Universities for Research in Astronomy, Inc., under NASA contract NAS5-26555.}

\begin{abstract}
We present the correlations between stellar mass, star formation rate (SFR) and [N~II]/H$\alpha$ flux ratio as indicator of gas-phase metallicity for a sample of 222 galaxies at $0.8 < z < 2.6$ and $\log(M_*/\mathrm{M}_\odot)=9.0-11.5$ from the LUCI, SINS/zC-SINF and KMOS$^{\mathrm{3D}}$ surveys. This sample provides a unique analysis of the mass-metallicity relation (MZR) over an extended redshift range using consistent data analysis techniques and strong-line metallicity indicator. We find a constant slope at the low-mass end of the relation and can fully describe its redshift evolution through the evolution of the characteristic turnover mass where the relation begins to flatten at the asymptotic metallicity. 
At fixed mass and redshift, our data do not show a correlation between the [N~II]/H$\alpha$ ratio and SFR, which disagrees with the 0.2-0.3~dex offset in [N~II]/H$\alpha$ predicted by the ``fundamental relation'' between stellar mass, SFR and metallicity discussed in recent literature. However, the overall evolution towards lower [N~II]/H$\alpha$ at earlier times does broadly agree with these predictions. 

\subjectheadings{galaxies: high-redshift, galaxies: evolution, infrared: galaxies} 
\end{abstract}

\section{Introduction}
\label{sec:intro}

Observed relations between a galaxy's stellar mass, star formation rate (SFR), and gas-phase metallicity can provide crucial constraints for galaxy evolution models which aim to understand the build up of galaxies over cosmic time. The existence of a correlation between stellar mass and metallicity has been firmly established both locally \citep[e.g.,][]{Lequeux1979, Tremonti2004}, and out to $z=3.5$ \citep[e.g.,][]{Erb2006, me2012, Yuan2013, Zahid2013, Henry2013, Cullen2014, Steidel2014}. Systematic uncertainties in the derivation of gas-phase metallicities from emission-line diagnostics significantly influence the absolute normalisation and slope of the mass-metallicity relation \citep[MZR;][]{Kewley2008}, and have complicated measurements of its evolution with redshift.
Recently, it has been found that including SFR as a secondary parameter in the correlation greatly reduces the scatter in the local MZR \citep{Ellison2008, Mannucci2010, Laralopez2010, Yates2012, Andrews2013}, though much less so at high redshift \citep[e.g.][]{Zahid2013,Steidel2014}.
\cite{Mannucci2010} have proposed a fundamental metallicity relation between galaxy abundance, mass and SFR that does not evolve with redshift out to $z=2.5$. 
Theoretical studies ranging from analytic equilibrium models \citep[e.g.,][]{Lilly2013} to cosmological hydrodynamical simulations \citep[e.g.,][]{Dave2011,Hirschmann2013} offer their own predictions of the MZR and the correlation with SFR; comparisons with observations provide crucial constraints on the physical processes and assumptions included in these models.

At high redshift, metallicity studies have long been based on relatively small samples due to the inherent difficulties of near-IR spectroscopy. They furthermore often remain limited to a narrow redshift range, such that attempts to constrain redshift evolution of the mass-metallicity relation necessarily rely on a comparison of different sample selections and diagnostics at each epoch. 
In this paper, we present a sample of 222 galaxies over a wide redshift range $0.8<z<2.6$ and mass range $\log(M_*/\mathrm{M}_\odot)=9.0-11.5$, for which we observed the H$\alpha$ and [N~II] emission with a combination of the multi-object LUCI spectrograph at the Large Binocular Telescope (LBT) in Arizona, and the SINFONI and KMOS integral field (IFU) instruments at the Very Large Telescope (VLT) in Chile. The large size and extended redshift coverage of this sample allows a consistent analysis of the correlations between stellar mass, SFR and gas-phase metallicity as traced by [N~II]/H$\alpha$, as well as their cosmic evolution.
We adopt the \cite{Chabrier2003} initial mass function and a flat cosmology with $\Omega_M = 0.3$ and H$_0 = 70$\,km\,s$^{-1}$\,Mpc$^{-1}$.



\begin{figure*}
\centering
\includegraphics[width=\textwidth]{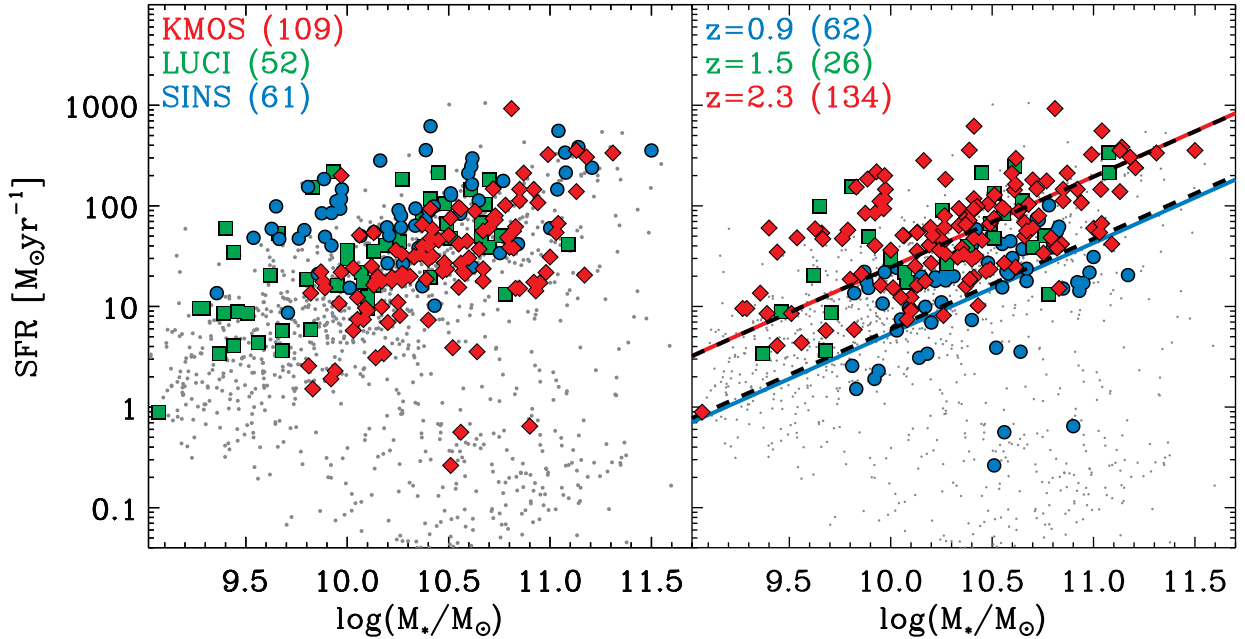}
\caption{Location of our sample in the SFR - stellar mass diagram. We have color-coded the data points by instrument in the \textit{left} panel and redshift in the \textit{right} panel. The numbers in parentheses note the number of sources in each survey or redshift bin. As a reference, the small gray dots represent the mass-selected galaxy population at $0.7 < z < 2.7$ in the CANDELS/3D-HST fields. In the \textit{right} panel, the blue and red solid lines show the star formation main-sequence from \cite{Lilly2013}  at $z=0.9$ and $z=2.3$ respectively. The dashed black lines represent the best fit relations to our spectroscopic sample at these redshifts. \label{fig:sfrm}}
\end{figure*}


\section{Observations and Data Reduction}
\label{sec:data}
\subsection{KMOS}
\label{subsec:data_kmos}
The KMOS$^{\mathrm{3D}}$ GTO survey is targeting a mass-selected sample of star-forming galaxies (SFG) at $0.7<z<2.7$ in the COSMOS, GOODS-South and UDS deep fields with the KMOS multi-object integral field spectrograph at the VLT \citep{Sharples2013}. Publicly available optical spectroscopic redshifts in these fields are supplemented with grism redshifts of all sources with $H_{140,\mathrm{AB}} < 24$~mag from the 3D-HST Treasury Survey \citep{Brammer2012}. This results in a target sample with reduced bias towards blue, star-forming, dust-free galaxies as inherent to samples based solely on optical spectroscopic redshift. The galaxy sample studied in this paper originates from commissioning data and the first semester of GTO observations, during which 105 unique targets at $<z>=0.9$ and 67 targets at $<z>=2.3$ were observed with median on-source integration times of 4h and 8.3h respectively \citep{Wisnioski2014}. The data were reduced using the Software Package for Astronomical Reductions with KMOS \citep[SPARK; ][]{Davies2013}.

The KMOS data cubes are analysed with the custom tool \textsc{linefit}, following procedures described in detail by \cite{Forster2009}. We smooth each cube with a 3 pixel-wide filter along the spatial axes. When the H$\alpha$ emission line is detected at $S/N>3$ in an individual spatial pixel, we proceed to fit H$\alpha$ and [N~II] simultaneously, forcing a common line width and redshift and locking the flux ratio of the [N~II] doublet to its theoretical value of 3.071 \citep{Storey2000}. With current integration times, we are able to derive a robust velocity field for 62/107 targets at $<z>=0.9$ and 47/67 targets at $<z>=2.3$.

\subsection{LUCI}
\label{subsec:data_luci}
From December 2009 to May 2012, we observed 148 SFGs at $1.3 < z < 2.5$ in the GOODS-North deep field and 8 SFGs at $z\sim2.3$ in the Q2343 field \citep{Steidel2004} with the multi-slit near-IR spectrograph LUCI at the LBT \citep{Seifert2010} for a median integration time of 4~hours. Targets were purely selected on spectroscopic redshift, based on version 1.0 of the GOODS-N PEP multi-wavelength catalog \citep{Berta2011}. We employed 1\arcsec\ wide slits with the 210zJHK and 150Ks gratings, which provide a spectral resolution $R\sim2900$ and $R\sim1900$ respectively. The observations were reduced employing a custom pipeline developed at MPE, which includes bad pixel masking, cosmic ray removal, distortion correction and optimal sky subtraction based on \cite{Davies2007}. For this work, we include 52 SFGs where H$\alpha$ is detected at $S/N>3$, and the H$\alpha$ and [N~II] lines are not contaminated by skylines.

\subsection{SINS/zC-SINF}
\label{subsec:data_sins}
The SINS and zC-SINF surveys have observed $>100$ SFGs at $1.3 < z < 2.5$ with the near-IR IFU spectrograph SINFONI at the VLT \citep{Eisenhauer2003}, with on-source integration times ranging from 1h to more than 20h. Target selection, observations and data reduction are described in detail by \cite{Forster2009} and \cite{Mancini2011}. Targets were selected from optical spectroscopic surveys of various parent samples photometrically selected based on rest-frame UV/optical magnitudes or colors. 35 galaxies have been followed up with adaptive optics (F\"{o}rster Schreiber et al. 2014, in prep.). Here we use a subsample of 61 galaxies with robust kinematics; they are detected with S/N=$23\pm12$ in H$\alpha$.

\section{Methodology}
\label{sec:methods}
\subsection{SFR, stellar mass and the [N~II]/H$\alpha$ line ratio}
\label{subsec:met}
We derive the SFRs and stellar masses for our sample from a combination of 3D-HST grism spectroscopy \citep{Brammer2012} with multi-wavelength rest-frame UV/optical photometry \citep{Skelton2014} and far-IR photometry \citep[available for 55\% of our sample;][]{Lutz2011, Magnelli2013}, using standard spectral energy distribution fitting techniques and a ladder of SFR indicators \citep{Wuyts2011}. Our analysis of the gas-phase metallicities is carried out as much as possible in terms of the directly observable [N~II]/H$\alpha$ line  ratio to avoid systematics associated with the choice of strong-line metallicity calibration \citep{Kewley2008}. When necessary, we employ the linear metallicity calibration by \cite{pp04} to relate [N~II]/H$\alpha$ to the oxygen abundance $12+\log(\mathrm{O/H})$\footnotemark[1], which has an intrinsic dispersion of 0.18~dex.
\footnotetext[1]{$12+\mathrm{\log(O/H)} = 8.9+0.57 \times \mathrm{\log([N~II]/H\alpha)}$}

For each KMOS and SINFONI data cube, we measure a global [N~II]/H$\alpha$ ratio from the galaxy-integrated 1D spectrum within a maximal elliptical aperture positioned at the galaxy center, tilted along the kinematic position angle, and with an ellipticity that matches the outer H$\alpha$ isophotes. Within this aperture, the spectra of individual spatial pixels are co-added after being velocity-shifted to a common H$\alpha$ centroid based on the galaxy's velocity field. We detect [N~II] at S/N$>2$ for 61/62 KMOS targets at $<z>=0.9$ and 41/47 targets at $<z>=2.3$. For SINFONI, [N~II] is detected at $S/N>2$ for 10/12 targets at $<z>=1.55$ and 40/49 targets at $<z>=2.3$. The long-slit LUCI spectra are not corrected for the velocity structure of the source; we attempted this correction for a subset of the best-sampled targets and did not find a significant effect on the emission line ratios.
We measure [N~II] at S/N$>2$ for 12/14 LUCI targets at $<z>=1.45$ and 20/38 LUCI targets at $<z>=2.3$. For the 17\% non-detections in our combined sample, we define 2$\sigma$ upper limits on [N~II] from the noise at the expected line position and the common linewidth.

\subsection{AGN contamination}
We note that using [N~II] emission as a metallicity indicator comes with its own complications, such as variations in the N/O ratio \citep{Kennicutt2003}, and saturation at high metallicities \citep{Kewley2002}. An important issue is the enhancement of [N~II]/H$\alpha$ for galaxies where an active galactic nucleus contributes to the ionizing radiation or shocks affect the ionization balance \citep{Kewley2013, Newman2014}. We identify 18 AGN in our sample from X-ray and radio data, infrared colors and rest-UV spectroscopy \citep[see][for more details]{Genzel2014}. Additionally, recent IFU studies have found evidence for an AGN in the central regions of massive $z=1-2$ SFGs from broad outflow components and/or enhanced line ratios \citep{Wright2010, Forster2014, Newman2014}. \cite{Genzel2014} report broad (FWHM$\gtrsim$1000~km/s) emission components associated with the nuclear regions of 20 targets included in our combined sample.

\subsection{Stacking}
\begin{figure*}
\centering
\includegraphics[width=\textwidth]{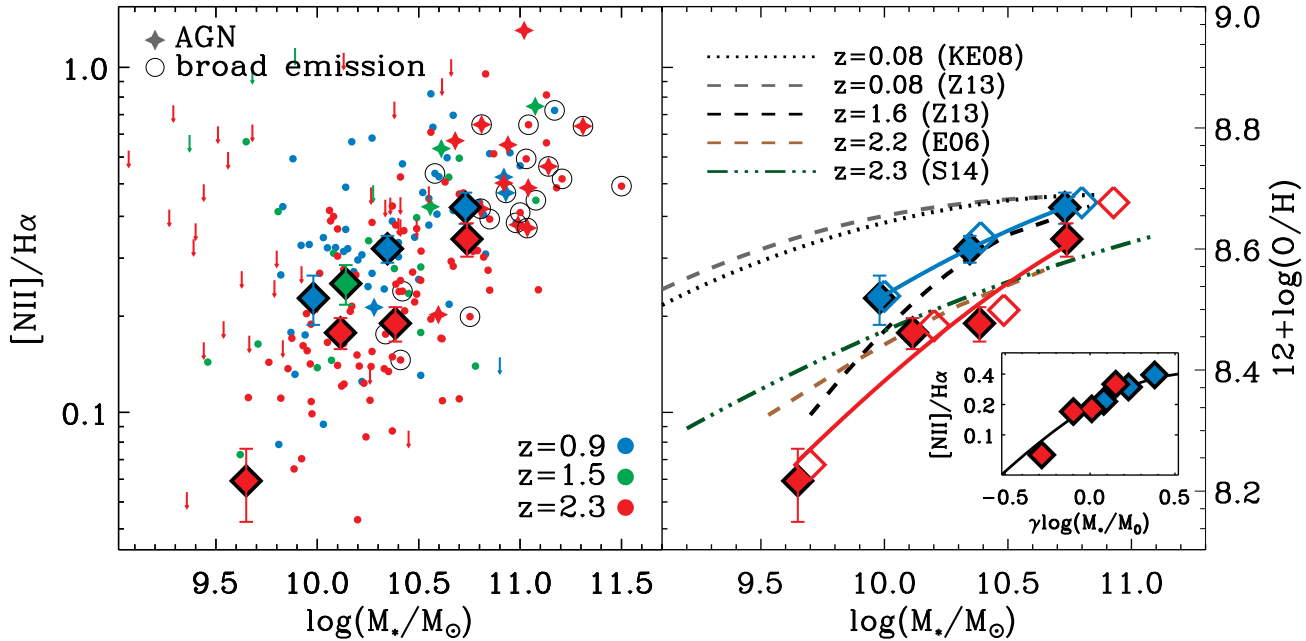}
\caption{[N~II]/H$\alpha$ flux ratio versus stellar mass for our combined sample, color-coded by redshift. The \textit{left} panel shows individual detections and 2$\sigma$ upper limits. As explained in the text, AGN identified from classic X-ray etc.\ indicators and broad nuclear AGN-driven outflows are indicated with four-pointed stars and black circles respectively. Large colored diamonds indicate the stacked [N~II]/H$\alpha$ ratios in three, one and four bins of stellar mass for the $z=0.9$,  $z=1.5$ and $z=2.3$ redshift slices, excluding the AGN-contaminated sources. The blue and red solid lines in the \textit{right} panel correspond to our best-fit MZRs as parametrised in Table~\ref{tab}. For reference, the open diamonds show the stacked results when all galaxies are included. We include the local MZR from \cite{Zahid2013} and \cite{Kewley2008}, as well as literature relations at $z=1.6$ \citep{Zahid2013}, $z=2.2$ \citep{Erb2006}, and $z=2.3$ \citep{Steidel2014}. The inset shows the stacked [N~II]/H$\alpha$ ratios at $z=0.9$ and $z=2.3$ as a function of $\gamma \log(M_*/M_0)$ with fixed local slope $\gamma = 0.4$. \label{fig:massmet}}. 
\end{figure*}

To include the [N~II] upper limits in our analysis, we stack the galaxies in bins of stellar mass. The 22 galaxies at $z=1.5$ do not suffice for an analysis as a function of stellar mass and will therefore not be discussed further, though we note that stacking all $z=1.5$ sources together results in an [N~II]/H$\alpha$ ratio intermediate between the $z=0.9$ and $z=2.3$ samples and consistent with previous literature results (Figure~\ref{fig:massmet}). At $z=0.9$ and $z=2.3$, we stack galaxies in 3 or 4 mass bins respectively, after removing the AGN and galaxies with broad emission components from our sample. For each mass bin, the 1D source spectra are continuum-subtracted, de-redshifted and normalised by the total H$\alpha$ flux. We exclude wavelength regions badly affected by skylines and robustly derive emission line fluxes and uncertainties for the stacked spectra as the jack-knife mean and standard error. The median stellar mass and SFR, and the stacked [N~II]/H$\alpha$ ratio for each bin is reported in Table~\ref{tab}. At $z=2.3$ we checked that stacking each of the KMOS, LUCI and SINS subsamples separately gives consistent results.

\section{Results}
\label{sec:results}

\subsection{The Mass-Metallicity Relation}
\label{subsec:massmet}
Figure~\ref{fig:sfrm} situates our sample in the SFR - stellar mass diagram and shows consistency with the star formation main sequence as parametrised by \cite{Lilly2013} at $z=0.9$ and $z=2.3$. In Figure~\ref{fig:massmet} we show the galaxy-integrated [N~II]/H$\alpha$ ratios for all sources as a function of their stellar mass. Symbols are color-coded by redshift, the mean uncertainty in [N~II]/H$\alpha$ is $\pm0.05$ (ranging from 0.039 to 0.066 for the various bands and surveys). AGNs are shown with four-pointed stars and galaxies with broad outflow components are identified with surrounding black circles. The stacked results (which exclude the AGNs and broad emission galaxies) are shown with filled colored diamonds. For reference, we obtain consistent results when including all galaxies in the stacks, as shown with the open diamonds in the right panel. Thus, while it remains important to check, the $\sim10$\% AGN contamination of our sample does not have a significant effect on the derived MZRs.

For comparison, the local MZR from \cite{Zahid2013} based on [N~II]/H$\alpha$ is included, which agrees with the result from \cite{Kewley2008} once this is converted to the linear instead of the cubic metallicity calibration from \cite{pp04}. At high redshift, we compare to [N~II]/H$\alpha$ based relations at $z=1.6$ \citep{Zahid2013}, $z=2.2$ \citep{Erb2006}\footnotemark[2] and $z=2.3$ \citep{Steidel2014}.
\footnotetext[2]{We use the \textit{current} stellar masses re-derived by \cite{Zahid2013} instead of the \textit{total} stellar masses reported by \cite{Erb2006}} 
We find good agreement towards high stellar masses, and some variation in the slope at the low mass end. This is likely due to differences in sample selection. \cite{Juneau2014} have recently reported a steeper local MZR after applying an H$\alpha$ luminosity threshold to a sample of SDSS galaxies. A common problem is the bias of high-z spectroscopic studies against red, dusty objects, which given the positive correlation between dust extinction and metallicity \cite[e.g.,][]{Zahid2013}, are generally metal-rich. 
A detailed comparison of the selection of the various high-z samples, including their rest-frame $U-V$ colours as a measure of extinction, would be very instructive for a better understanding of the varying slopes. 
\begin{table*}
\centering
\caption{Mass-metallicity relation \label{tab}}

\medskip

\begin{tabularx}{\linewidth}{*{5}{p{.2\linewidth}}}
\multicolumn{5}{c}{Properties for each stellar mass bin} \\ \toprule 
Redshift  & $\log(M_*/\mathrm{M}_\odot)$ & SFR & [N~II]/H$\alpha$ & \#targets \\ \midrule
\multirow{3}{*}{z=0.9}      & $9.98_{-0.12}^{+0.16}$    & $14_{-11}^{+8}$    & $0.21\pm0.03$   &  18  \\ 
                                        & $10.35_{-0.13}^{+0.16}$  & $19_{-13}^{+21}$  & $0.30\pm0.03$   &  19  \\
                                        & $10.73_{-0.14}^{+0.17}$  & $30_{-15}^{+32}$  & $0.39\pm0.04$   &  19  \\
\vspace{0.1cm} \\
\multirow{3}{*}{z=2.3}      & $9.65_{-0.16}^{+0.27}$    & $40_{-34}^{+71}$  & $0.06\pm0.02$   &  28  \\
                                        & $10.11_{-0.11}^{+0.12}$  & $41_{-28}^{+76}$   & $0.17\pm0.02$   &  28  \\
                                        & $10.39_{-0.06}^{+0.08}$  & $64_{-41}^{+41}$  & $0.18\pm0.02$   &  28  \\
                                        & $10.74_{-0.18}^{+0.13}$  & $88_{-47}^{+93}$  & $0.32\pm0.04$   &  29  \\ \bottomrule
\end{tabularx}

\bigskip

\begin{tabularx}{\linewidth}{*{6}{p{.167\linewidth}}}
\multicolumn{6}{c}{Best-fit parameters} \\ \toprule
Reference & Redshift &  $Z_0$ & $\log(M_0 /\mathrm{M}_\odot)$ & $\gamma$ & $\log(M_0^\mathrm{fixed} /\mathrm{M}_\odot)$ \\ \midrule
Z13b       & 0.08   & $8.69\pm0.01$     & $9.02\pm0.02$  & $0.40\pm0.01$   &   $8.95\pm0.05$  \\ 
this work & 0.9     & $8.8\pm0.4$         & $10.2\pm0.9$    & $0.4\pm0.6$       &   $9.78\pm0.11$  \\
this work & 2.3     & $8.7\pm0.3$         & $10.5\pm0.5$    & $0.5\pm0.2$       &   $10.36\pm0.06$  \\
\end{tabularx}
\end{table*} 
   
\begin{figure*}
\centering
\includegraphics[width=\textwidth]{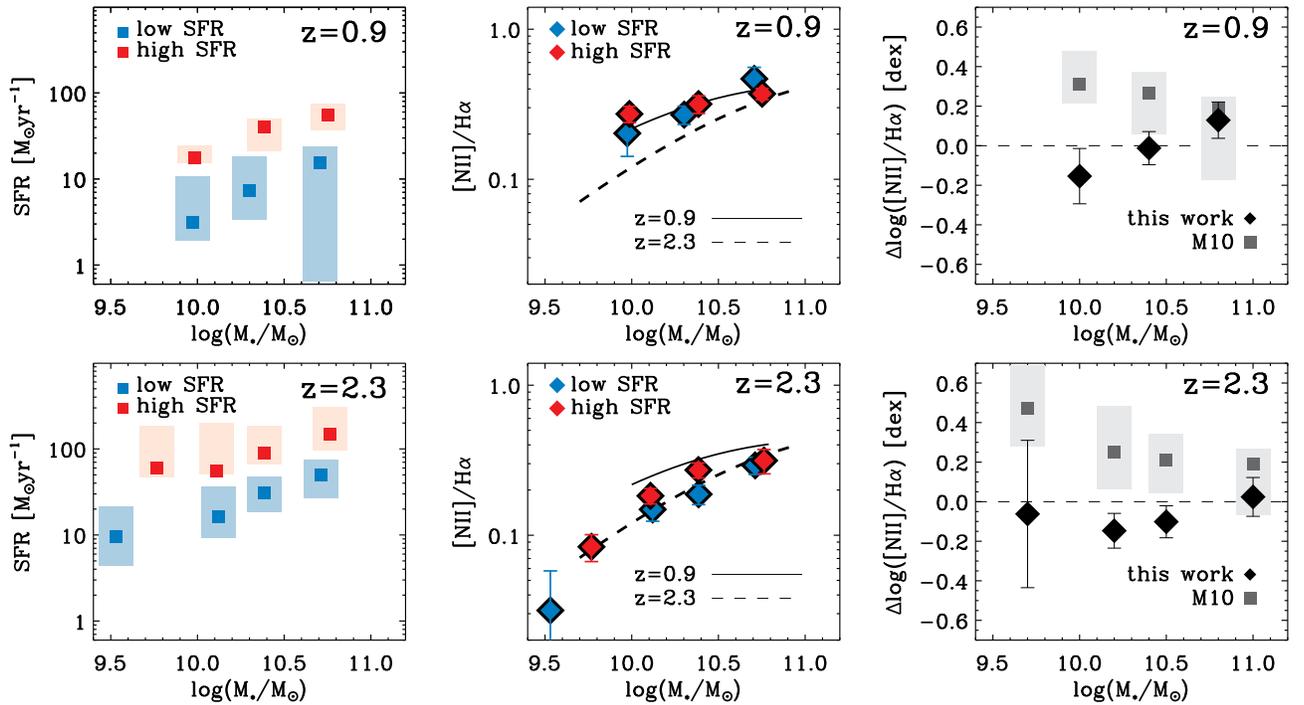}
\caption{We divide the sample at $z=0.9$ \textit{(top row)} and $z=2.3$ \textit{(bottom row)} into two SFR bins for each stellar mass bin. In the \textit{left} column, the shaded regions show the 1$\sigma$ dynamic range in SFR probed in each bin, the filled square corresponds to the median value. In the \textit{middle} column, the MZR for the low and high SFR bin is shown with blue and red diamonds respectively. The black solid and dashed lines correspond to our best-fit MZRs at $z=0.9$ and $z=2.3$ as shown in Figure~\ref{fig:massmet} and parametrised in Table~\ref{tab}. The MZR for each SFR bin is consistent with the best-fit MZR for the full sample at that redshift. Hence, at fixed mass and redshift we do not find a correlation between the [N~II]/H$\alpha$ flux ratio and SFR. The \textit{right} column shows the relative offset in $\log(\mathrm{[N~II]/H}\alpha)$ between SFR bins $\Delta \log(\mathrm{[N~II]/H}\alpha) = \log(\mathrm{[N~II]/H}\alpha)_{\mathrm{lowSFR}} - \log(\mathrm{[N~II]/H}\alpha)_{\mathrm{highSFR}}$. We compare our data (black diamonds) to the expectation from \cite{Mannucci2010} for the 1$\sigma$ dynamic range in SFR for each bin (shaded regions). \label{fig:sfr}}
\end{figure*}
\npar
We provide fits to the MZR based on the parametrisation introduced by \cite{Zahid2014} 
\begin{equation}
\label{eq1}
12+\mathrm{\log(O/H)} = Z_0 + \log \left[ 1 - \exp \left( - \left[ \frac{M_*}{M_0} \right] ^ \gamma \right) \right]
\end{equation}
where $Z_0$ corresponds to the asymptotic metallicity, $M_0$ is the characteristic turnover mass where the relation begins to flatten and $\gamma$ is the power-law slope at stellar masses $\ll M_0$. Out to $z=1.6$, \cite{Zahid2014} found a constant asymptotic metallicity and slope and a power-law increase of $M_0$ with redshift, such that redshift evolution of the MZR depends solely on the evolution of the characteristic turnover mass. They suggest this follows from the more fundamental universal relation between metallicity and stellar-to-gas mass ratio. Our best-fit parameters in Table~\ref{tab} confirm this result within the uncertainties out to $z=2.3$. The relation between metallicity and stellar mass scaled by $M_0$ is therefore independent of redshift, as can be seen in the inset in the right panel of Figure~\ref{fig:massmet}. 
The rightmost column in Table~\ref{tab} reports the best-fit $M_0^\mathrm{fixed}$ derived for a fixed $Z_0 = 8.69$ and $\gamma=0.40$ as found for the local relation. We can describe the redshift evolution of the characteristic turnover mass as 
\begin{equation}
\label{eq2}
\log(M_0/\mathrm{M}_\odot) = (8.86\pm0.05) + (2.92\pm0.16) \log(1+z)
\end{equation}    
This result is not strongly dependent on the choice of $Z_0$ and $\gamma$. \cite{Zahid2014} report a consistent slope for $M_0(z)$ within the uncertainties; the difference in zeropoint is likely due to metallicity calibration offsets between the R23 and N2 indicators.

\subsection{Star Formation Rate as a Secondary Parameter}
\label{subsec:fmr}

\begin{figure}
\centering
\includegraphics[width=9cm]{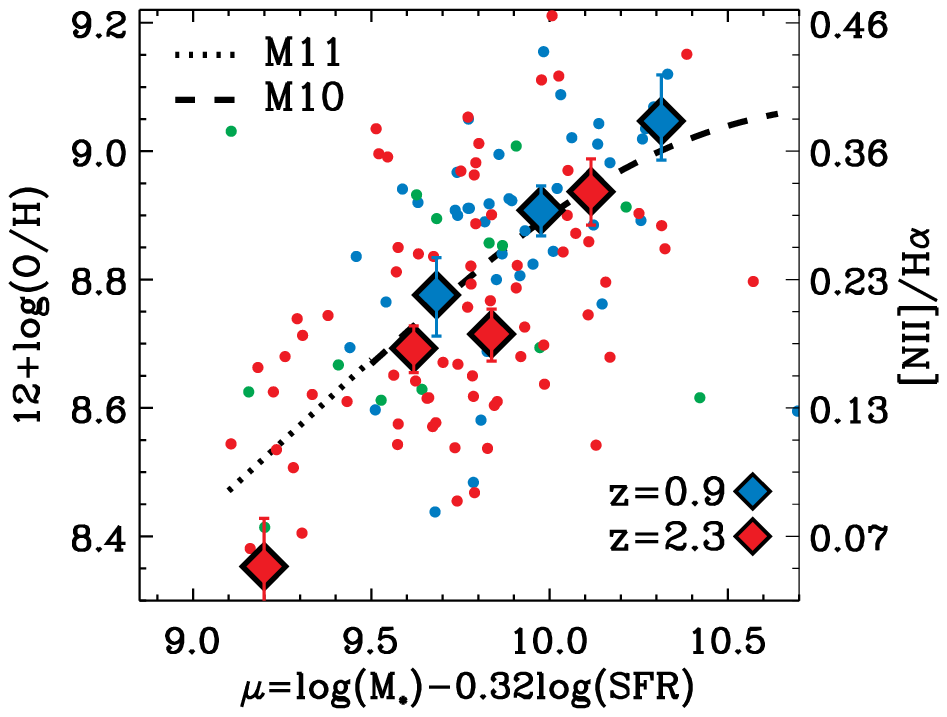}
\caption{Metallicity derived from the \cite{Maiolino2008} calibration of the [N~II]/H$\alpha$ flux ratio versus $\mu = \log(M_*) - 0.32 \log(SFR)$ for individual targets and spectra stacked in bins of $\mu$ at $z=0.9$ and $z=2.3$. For comparison, we show the local relation from \cite{Mannucci2010} \textit{(dashed line)} and the extension to lower stellar masses from \cite{Mannucci2011} \textit{(dotted line)}. \label{fig:fmr}}. 
\end{figure}

We investigate the role of SFR in the mass-metallicity relation by stacking our sample at $z=0.9$ and $z=2.3$ in two bins of SFR for each mass bin. The left column of Figure~\ref{fig:sfr} shows the 1$\sigma$ dynamic range in SFR probed by each bin, as well as the median value. As seen in the middle column, we do not find a correlation between the [N~II]/H$\alpha$ ratio and SFR at fixed mass and redshift. The MZR for the low and high SFR bins is consistent with the best-fit MZR derived for the complete sample at that redshift. This result is confirmed when we split each mass bin into three bins of SFR and compare the top and bottom third. 
In contrast, \cite{Zahid2013} do find a correlation for their sample at $z=1.6$.
The range of SFR probed is similar to our sample, but they use dust-corrected H$\alpha$-derived SFRs instead of the ladder of SFR$_{\mathrm{UV+IR}}$ used here. Using H$\alpha$ for both the SFR and metallicity measurement could introduce an artificial trend. Alternatively, the longer timescales probed by SFR$_{\mathrm{UV+IR}}$ might average out the correlation with gas-phase metallicity.

\npar
We compare our data to the fundamental relation between metallicity, stellar mass and SFR proposed by \cite{Mannucci2010}.  
Since their relation is based on the \cite{Maiolino2008} metallicity calibration, and our high-z sample probes significantly larger SFRs, one should be careful in interpreting a direct comparison. The shaded region in the right column of Figure~\ref{fig:sfr} shows the relative metallicity offset predicted by \cite{Mannucci2010} for the 1$\sigma$ dynamic range in SFR for our mass bins. The positive values for $\Delta \log(\mathrm{[N~II]/H}\alpha) = \log(\mathrm{[N~II]/H}\alpha)_{\mathrm{lowSFR}} - \log(\mathrm{[N~II]/H}\alpha)_{\mathrm{highSFR}}$ reflect the anti-correlation between SFR and metallicity they find for the SDSS sample, which becomes more pronounced at low stellar masses. In contrast, our data at fixed $z=0.9$ and $z=2.3$ show no relative offset in [N~II]/H$\alpha$ between SFR bins, or even a somewhat negative one. 
In Figure~\ref{fig:fmr} we directly plot metallicity versus $\mu = \log(M_*) - 0.32 \log(SFR)$, which has been proposed as the projection which minimizes the scatter in the local relation between metallicity, mass and SFR. To be consistent, here we use the \cite{Maiolino2008} calibration to derive metallicities from our [N~II]/H$\alpha$ ratios. We find an overall agreement with the relation proposed by \cite{Mannucci2010} and their extension towards lower stellar masses \citep{Mannucci2011}, though there is some tension for the lowest mass bin.

\section{Summary}
\label{sec:sum}
We report the [N~II]/H$\alpha$ flux ratios for a sample of 222 galaxies at $0.8 < z < 2.6$, probing a wide range of stellar mass $\log(M_*/\mathrm{M}_\odot)=9.0-11.5$. The extended redshift coverage allows the first consistent analysis of the evolution of the correlations between stellar mass, SFR and gas-phase metallicity with cosmic time. We detect [N~II] for 83\% of our sample and employ stacking techniques to extend the results down to $\log(M_*/\mathrm{M}_\odot)=10.0$ at $z=0.9$ and $\log(M_*/\mathrm{M}_\odot)=9.7$ at $z=2.3$. We find good agreement with other high-z MZRs in the literature, though a careful analysis of sample selection is necessary to interpret the slope towards low stellar masses. Our results at $z=0.9$ and $z=2.3$ show a common power-law slope with the local MZR within the uncertainties, such that the redshift evolution of the MZR can be fully determined by the evolution of the characteristic turnover mass $M_0$. 

In the context of the ``fundamental relations" between metallicity, stellar mass and SFR that have been found locally, the redshift evolution of the MZR towards lower abundances at earlier times has been interpreted as due to the higher SFRs of high-z SFGs. However, the lack of correlation between SFR and metallicity at fixed redshift and mass shown here, suggests that the redshift evolution of SFR and metallicity might not be causally related.


\begin{acknowledgments}
We are grateful to the referee for thoughtful comments which significantly improved the quality of this Letter. DJW and MF acknowledge the support of the Deutsche Forschungsgemeinschaft via Project ID 387/1-1.
\end{acknowledgments}




\begin{thebibliography}

\bibitem[{{Andrews} \& {Martini}(2013)}]{Andrews2013}
{Andrews}, B.~H., \& {Martini}, P. 2013, \apj, 765, 140


\bibitem[Berta et al.(2011)]{Berta2011} 
{Berta}, S. and {Magnelli}, B. and {Nordon}, R., et al. 2011, \aap, 532, A49

\bibitem[Brammer et al.(2012)]{Brammer2012} 
{Brammer}, G.~B. and {van Dokkum}, P.~G. and {Franx}, M., et al. 2012, \apjs, 200, 13


\bibitem[{{Chabrier}(2003)}]{Chabrier2003}
{Chabrier}, G. 2003, \pasp, 115, 763

\bibitem[Cullen et al.(2014)]{Cullen2014} 
{Cullen}, F., {Cirasuolo}, M., {McLure}, R.~J., {Dunlop}, J.~S., \& {Bowler}, R.~A.~A. 2014, \mnras, 440, 2300

\bibitem[Dav{\'e} et al.(2012)]{Dave2011} 	
{Dav{\'e}}, R., {Finlator}, K., \& {Oppenheimer}, B.~D. 2011, \mnras, 416, 1354


\bibitem[Davies(2007)]{Davies2007}
{Davies}, R.~I. 2007, \mnras, 375, 1099


\bibitem[Davies et al.(2013)]{Davies2013} 
{Davies}, R.~I., {Agudo Berbel}, A., {Wiezorrek}, E., et al. 2013, \aap, 558, A56

\bibitem[Eisenhauer et al.(2003)]{Eisenhauer2003} 
{Eisenhauer}, F., {Abuter}, R., {Bickert}, K., et al. 2003, Society of Photo-Optical Instrumentation Engineers (SPIE) Conference Series, 4841

\bibitem[Ellison et al.(2008)]{Ellison2008} 
{Ellison}, S.~L., {Patton}, D.~R., {Simard}, L., \& {McConnachie}, A.~W. 2008, \apjl, 672, L107 

\bibitem[Erb et al.(2006)]{Erb2006}
{Erb}, D.~K., {Shapley}, A.~E., {Pettini}, M., {Steidel}, C.~C., {Reddy}, N.~A., {Adelberger}, K.~L. 2006a, \apj, 644, 813

\bibitem[F\"{o}rster Schreiber et~al.(2009)]{Forster2009}
{F\"{o}rster Schreiber}, N.~M., Genzel, R., Bouch\'{e}, N., et al. 2009, \apj, 706, 1364

\bibitem[F\"{o}rster Schreiber et~al.(2014)]{Forster2014}
{F{\"o}rster Schreiber}, N.~M., {Genzel}, R., {Newman}, S.~F., et al. 2014, \apj, 787, 38

\bibitem[Genzel et al.(2014)]{Genzel2014} 
{Genzel}, R., {F{\"o}rster Schreiber}, N.~M., Rosario, D., et al. 2014, in prep.

\bibitem[Henry et al.(2013)]{Henry2013} 
{Henry}, A., {Scarlata}, C., {Dom{\'{\i}}nguez}, A., et al. 2013, \apjl, 776, L27

\bibitem[Hirschmann et al.(2013)]{Hirschmann2013} 
{Hirschmann}, M., {Naab}, T., {Dav{\'e}}, R, et al. 2013, \mnras, 436, 2929


\bibitem[Juneau et al.(2014)]{Juneau2014} 
Juneau, S., Bournaud, F., Charlot, S., et al. 2014, arXiv:1403.6832

\bibitem[Kennicutt et al.(2003)]{Kennicutt2003} 
{Kennicutt}, Jr., R.~C., {Bresolin}, F., \& {Garnett}, D.~R. 2003, \apj, 591, 801

\bibitem[Kewley \& Dopita(2002)]{Kewley2002} 
{Kewley}, L.~J., \& {Dopita}, M.~A. 2002, \apjs, 142, 35

\bibitem[Kewley \& Ellison(2008)]{Kewley2008} 
{Kewley}, L.~J. \& {Ellison}, S.~L. 2008, \apj, 681, 1183

\bibitem[Kewley et al.(2013)]{Kewley2013} 
{Kewley}, L.~J., {Dopita}, M.~A., {Leitherer}, C., et al. 2013, \apj, 774, 100

\bibitem[Lara-L{\'o}pez et al.(2010)]{Laralopez2010} 
{Lara-L{\'o}pez}, M.~A., Cepa, J., Bongiovanni, A., et al. 2010, \aap, 521, L53


\bibitem[Lequeux et al.(1979)]{Lequeux1979} 
{Lequeux}, J., {Peimbert}, M., {Rayo}, J.~F., {Serrano}, A., \& {Torres-Peimbert}, S. 1979, \aap, 80, 155

\bibitem[Lilly et al.(2013)]{Lilly2013} 
{Lilly}, S.~J., {Carollo}, C.~M., {Pipino}, A., {Renzini}, A., \& {Peng}, Y. 2013, \apj, 772, 119
	

\bibitem[Lutz et al.(2011)]{Lutz2011} 
{Lutz}, D., {Poglitsch}, A., {Altieri}, B., et al. 2011, \aap, 532, A90

\bibitem[Maiolino et al.(2008)]{Maiolino2008} 
{Maiolino}, R., Nagao, T., Grazian, A., et al. 2008, \aap, 488, 463

\bibitem[Magnelli et al.(2013)]{Magnelli2013} 
{Magnelli}, B., {Popesso}, P., {Berta}, S., et al. 2013, \aap, 553, A132

\bibitem[Mancini et al.(2011)]{Mancini2011} 
{Mancini}, C., {F{\"o}rster Schreiber}, N.~M., {Renzini}, A., et al. 2011, \apj, 743, 86


\bibitem[Mannucci et al.(2010)]{Mannucci2010} 
{Mannucci}, F., {Cresci}, G., {Maiolino}, R., {Marconi}, A., \&	{Gnerucci}, A. 2010, \mnras, 408, 2115

\bibitem[Mannucci et al.(2011)]{Mannucci2011} 
{Mannucci}, F., {Salvaterra}, R., \& {Campisi}, M.~A. 2011, \mnras, 414, 1263

\bibitem[Markwardt(2009)]{mpfitfun} 
Markwardt, C.~B.\ 2009, Astronomical Society of the Pacific Conference Series, 411, 251 



\bibitem[Newman et al.(2014)]{Newman2014} 
Newman, S.~F.,  Buschkamp, P., Genzel, R., et al. 2014, \apj, 781, 21

\bibitem[{{Pettini} \& {Pagel}(2004)}]{pp04}
{Pettini}, M. \& {Pagel}, B.~E.~J. 2004, \mnras, 348, L59


\bibitem[S{\'a}nchez et al.(2013)]{Sanchez2013}
{S{\'a}nchez}, S.~F., {Rosales-Ortega}, F.~F., {Jungwiert}, B., et al. 2013, \aap, 554, A58


\bibitem[Seifert et al.(2010)]{Seifert2010}
{Seifert}, W., {Ageorges}, N., {Lehmitz}, M., et al. 2010, Society of Photo-Optical Instrumentation Engineers (SPIE) Conference Series, 7735




\bibitem[Sharples et al.(2013)]{Sharples2013}
{Sharples}, R.,  {Bender}, R.,  {Agudo Berbel}, A. et al. 2013, The Messenger, 151, 21

\bibitem[Skelton et al.(2014)]{Skelton2014}
{Skelton}, R.~E. and {Whitaker}, K.~E. and {Momcheva}, I.~G., et al. 2014, arXiv:1403.3689

\bibitem[Steidel et al.(2004)]{Steidel2004}
{Steidel}, C.~C., {Shapley}, A.~E., {Pettini}, M., {Adelberger}, K.~L., {Erb}, D.~K., {Reddy}, N.~A., \& {Hunt}, M.~P. 2004, \apj, 604, 534

\bibitem[Steidel et al.(2014)]{Steidel2014}
{Steidel}, C.~C., Rudie, G.~C., Strom, A.~L., et al. 2014, arXiv:1405.5473
	
\bibitem[Storey \& Zeippen(2000)]{Storey2000} 
Storey, P.~J., \& Zeippen, C.~J.\ 2000, \mnras, 312, 813 


\bibitem[Tremonti et al.(2004)]{Tremonti2004}
{Tremonti}, C.~A., Heckman, T.~M., Kauffmann, G., et al. 2004, \apj, 613, 898


\bibitem[van Dokkum et al.(2013)]{vandokkum2013}
{van Dokkum}, P., {Brammer}, G., {Momcheva}, I., {Skelton}, R.~E., {Whitaker}, K.~E. and {for the 3D-HST team} 2013, arXiv:1305.2140


\bibitem[Wisnioski et al.(2014)]{Wisnioski2014}
{Wisnioski}, E., {F\"{o}rster Schreiber}, N.~M., Wuyts, E., et al. 2014, in prep.

\bibitem[Wright et al.(2010)]{Wright2010}
{Wright}, S.~A., {Larkin}, J.~E., {Graham}, J.~R., \& {Ma}, C.-P. 2010, \apj, 711, 1291
	
\bibitem[Wuyts, S. et al.(2011)]{Wuyts2011}
{Wuyts}, S. and {F{\"o}rster Schreiber}, N.~M. and {Lutz}, D., et al. 2011, \apj, 738, 106

\bibitem[Wuyts, E. et al.(2012)]{me2012}
{Wuyts}, E., {Rigby}, J.~R., {Sharon}, K., \& {Gladders}, M.~D. 2012, \apj, 755, 73


	
\bibitem[Yates et al.(2012)]{Yates2012}
{Yates}, R.~M., {Kauffmann}, G., \& {Guo}, Q. 2012, \mnras, 422, 215
	
\bibitem[{Yuan} {et~al.}(2013)]{Yuan2013}
{Yuan}, T.-T., {Kewley}, L.~J., \& {Rich}, J. 2013, \apj, 767, 106


\bibitem[Zahid et al.(2013)]{Zahid2013}
{Zahid}, H.~J., {Kashino}, D., {Silverman}, J.~D., et al. 2013, arXiv:1310.4950

\bibitem[Zahid et al.(2014)]{Zahid2014}
{Zahid}, J., {Dima}, G., {Kudritzki}, R., {Kewley}, L., {Geller}, M., {Hwang}, H.~S. 2014, arXiv:1404.7526

\end{thebibliography}
\end{document}